\documentclass[final,3p,times,twocolumn]{elsarticle}
\usepackage{graphicx}
\usepackage{amssymb}

\journal{Physics Letters B}
\begin{document}
\begin{frontmatter}
\title{Isospin properties of electric dipole excitations in $^{48}$Ca}

\author[ikp]{V.~Derya\corref{cor1}}
\ead{derya@ikp.uni-koeln.de}
\address[ikp]{Institut f\"ur Kernphysik, Universit\"at zu K\"oln, 50937 K\"oln, Germany}

\author[emmi,fias]{D.~Savran}
\address[emmi]{ExtreMe Matter Institute EMMI and Research Division, GSI, 64291 Darmstadt, Germany}
\address[fias]{Frankfurt Institute for Advanced Studies FIAS, 60438 Frankfurt am Main, Germany}

\author[ikp]{J.~Endres}

\author[kvi,ganil]{M.~N.~Harakeh}
\address[kvi]{Kernfysisch Versneller Instituut, University of Groningen, 9747 AA Groningen, The Netherlands}
\address[ganil]{GANIL, CEA/DSM-CNRS/IN2P3, 14076 Caen, France}

\author[ohio]{H.~Hergert}
\address[ohio]{Department of Physics, Ohio State University, Columbus, OH 43210, USA}

\author[tunl,ncsu]{J.~H.~Kelley}
\address[tunl]{Triangle Universities Nuclear Laboratory, Durham, NC 27708, USA}
\address[ncsu]{Department of Physics, North Carolina State University, Raleigh, NC 27695, USA}

\author[risp]{P.~Papakonstantinou}
\address[risp]{Rare Isotope Science Project, Institute for Basic Science, Daejeon 305-811, Republic of Korea}

\author[darm]{N.~Pietralla}
\address[darm]{Institut f\"ur Kernphysik, TU Darmstadt, 64289 Darmstadt, Germany}

\author[darm]{V.~Yu.~Ponomarev}

\author[darm]{R.~Roth}

\author[duke,tunl]{G.~Rusev\fnref{losalamos}}
\fntext[losalamos]{Present address: Chemistry Division, Los Alamos National Laboratory, Los Alamos, NM 87545, USA.}

\author[duke,tunl]{A.~P.~Tonchev\fnref{LLNL}}
\fntext[LLNL]{Present address: Physics Division, Lawrence Livermore National Laboratory, Livermore, CA 94550, USA.}

\author[duke,tunl]{W.~Tornow}
\address[duke]{Department of Physics, Duke University, Durham, NC 27708, USA}

\author[kvi]{H.~J.~W\"ortche}

\author[ikp]{A.~Zilges}

\cortext[cor1]{Corresponding author}

\begin{abstract}
Two different experimental approaches were combined to study the electric dipole strength in the doubly-magic nucleus $^{48}$Ca below the neutron threshold. Real-photon scattering experiments using bremsstrahlung up to 9.9~MeV and nearly mono-energetic linearly polarized photons with energies between 6.6 and 9.51~MeV provided strength distribution and parities, and an $(\alpha,\alpha'\gamma)$ experiment at $E_{\alpha}=136$~MeV gave cross sections for an isoscalar probe. The unexpected difference observed in the dipole response is compared to calculations using the first-order random-phase approximation and points to an energy-dependent isospin character. A strong isoscalar state at 7.6~MeV was identified for the first time supporting a recent theoretical prediction.
\end{abstract}

\begin{keyword}

low-lying electric dipole excitations \sep $^{48}$Ca \sep isospin character

\end{keyword}

\end{frontmatter}

Doubly-magic nuclei are exceptional cases for studying nuclear-structure properties. On one hand, such nuclei are traditionally considered as key-stones for testing theoretical approaches, such as, for example, the random-phase approximation (RPA) \cite{Paar07}. On the other hand, the low level density allows a detailed experimental investigation of individual excitations using different probes, therefore giving access to a variety of observables which provide a well-understood basis to confront nuclear models with. For the investigation of the isospin dependence of excitations in the many-body quantum system of an atomic nucleus consisting of few to many nucleons in two different isospin states (proton and neutron), the calcium chain offers a unique case with two stable doubly-magic nuclei, the $N=Z$ nucleus $^{40}$Ca and the neutron-rich $^{48}$Ca with $N/Z=1.4$. Dominant isoscalar (IS) and isovector (IV) excitation modes at excitation energies above the particle thresholds are well known, as for example the IV giant dipole resonance (IVGDR) or the IS giant quadrupole resonance \cite{harakeh}. Recently, an unexpected evolution of low-lying IS electric dipole strength as a function of the neutron number was predicted in the Ca chain \cite{Papa12}. This transition of the structure of the excitations from proton-skin oscillation to pure IS oscillation to neutron-skin oscillation depends on the used interaction or, more generally, the theoretical approach. However, isospin-sensitive experiments can address this question.

In this Letter, we report on experimental results of electric dipole ($E1$) excitations in $^{48}$Ca using different experimental techniques, which revealed different properties of single excitations. Moreover, we compare these results directly to results from RPA calculations. Interest in low-lying $E1$ strength is based on the observation of concentrated, mostly bound, $J^{\pi}=1^-$ states in spherical medium-heavy and heavy neutron-rich nuclei. This so-called pygmy dipole resonance (PDR) \cite{Brzo69,Zilg02}, and low-lying $1^-$ states in general, have been investigated in experimental and theoretical studies (see Refs. \cite{Savr13,Paar07,Tani13} and references therein). In addition, the diversity and interplay of different low-lying $E1$ modes, including a significant IS toroidal mode \cite{Repk13}, are under recent discussion. A common method for probing dipole strength in stable nuclei below the neutron threshold is real-photon scattering \cite{Zilg02, Tonc10, Savr11, Schw13}. Additional experimental techniques applied include, e.g., Coulomb excitation using protons \cite{Tami11} and $(\alpha,\alpha'\gamma)$ coincidence experiments dominated by the strong interaction \cite{Savr06b}. Inelastic-scattering experiments in inverse kinematics using radioactive beams, as well as Coulomb dissociation and Coulomb excitation experiments using radioactive beams, can provide insight into dipole strength distributions of very neutron-proton asymmetric nuclei \cite{Adri05, Klim07a, Wiel09}. One of the remaining open questions is the systematic evolution and nature of $E1$ excitations for nuclei with different masses and neutron-to-proton ratios. In this regard, Ca isotopes are of particular interest because of their light-to-medium mass and their wide range of $N/Z$ ratios.                                                                                                                                                                        

Low-lying $E1$ excitations in the calcium chain have been systematically investigated by real-photon scattering experiments \cite{Hart00,Hart02,Hart04,Isaa11} performed at the Darmstadt High-Intensity Photon Setup \cite{Sonn11} at the S-DALINAC. The experiments showed that the exhaustion of the Thomas-Reiche-Kuhn (TRK) energy-weighted sum rule (EWSR) for IV $E1$ transitions \cite{harakeh} up to 10~MeV increases from 0.020(3)\% for $^{40}$Ca to 0.39(4)\% for $^{48}$Ca. To complete the $N/Z$ systematics, a measurement was conducted on the nucleus in between, $^{44}$Ca, which exhausts 0.39(7)\% of the EWSR. Microscopic calculations using the extended theory of finite Fermi systems \cite{Kame04} reproduce the trend of the EWSR evolution of $E1$ strength along the Ca chain \cite{Tert07a}. However, during the last years, it became evident that one needs additional experimental probes to understand the structure of the $E1$ strength. In neutron-rich spherical nuclei in the medium-mass region the low-lying $E1$ strength was investigated in systematic studies by means of the $(\gamma,\gamma')$ and $(\alpha,\alpha'\gamma)$ reactions \cite{Savr06a,Savr06b,Endr09,Endr10,Endr12,Dery13}. These systematic studies revealed an isospin splitting into lower-energy isospin-mixed $E1$ excitations, usually referred to as the ``real" PDR, and higher-energy isovector-dominated $E1$ excitations, which belong to the tail of the IVGDR \cite{Tson08,Endr10}.

Driven by these experiments and the previously discussed results of $(\gamma,\gamma')$ experiments on the calcium isotopes, the present work employed the $(\alpha,\alpha'\gamma)$ reaction on $^{48}$Ca to learn about the nature of its $E1$ excitations, which was also discussed in a recent theoretical work \cite{Papa12}. The $\alpha$-$\gamma$ coincidence method, in which inelastically scattered $\alpha$ particles are measured in coincidence with the subsequently emitted $\gamma$ rays, was used \cite{Savr06a}. The $\alpha$-particle beam at an energy of 136~MeV and an average beam current of 1.0 particle nA was provided by the AGOR cyclotron at the Kernfysisch Versneller Instituut in Groningen, The Netherlands. The scattered $\alpha$ particles were detected by the EUROSUPERNOVA detection system \cite{Hage99} of the QQD-type Big-Bite Spectrometer (BBS) \cite{Berg95} at $\theta_{lab}=5.8^\circ$. The BBS solid-angle coverage was $\Delta\Omega_{\alpha}=9.2$~msr, with a horizontal opening angle of $4^\circ$, and the excitation-energy resolution in a singles $\alpha$-scattering measurement was 236(1)~keV at 3831~keV. For $\gamma$-ray spectroscopy, an array of six high-purity germanium (HPGe) detectors, each with an opening angle of around 20$^{\circ}$, was mounted around the target chamber and achieved an absolute photopeak-efficiency of 0.504(1)\% at 1238~keV. The self-supporting Ca target had a thickness of 1.7 mg/cm$^2$ and was enriched to 99\% with $^{48}$Ca.

\begin{figure}[htbp]
\centering
\includegraphics[width=8.3cm]{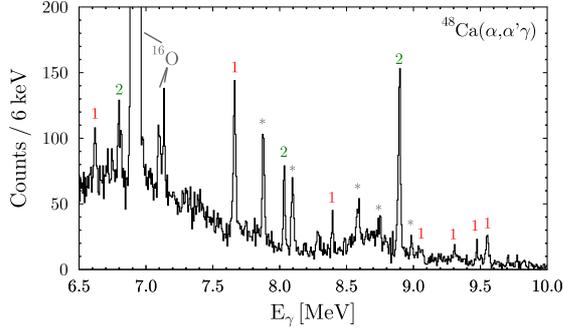}
\caption{Summed $\gamma$-ray spectrum of the HPGe detectors obtained by gating on de-exciting transitions to the ground state ($E_X\approx E_{\gamma}$), measured in the $^{48}$Ca$(\alpha,\alpha'\gamma)$ experiment. Observed electric transitions in $^{48}$Ca are labeled according to their multipolarity. Transitions in $^{40}$Ca are marked with stars, and transitions in $^{16}$O are labeled.}
\label{fig:GS_gamma}
\end{figure}

Previous experiments \cite{Savr06a,Savr06b,Endr09,Endr10,Dery13} demonstrated the advantages of the $\alpha$-$\gamma$ coincidence method, which allows us to select $\gamma$-ray transitions to the ground state or excited states. In this way an excellent selectivity to $E1$ excitations is achieved, providing a very good peak-to-background ratio in the projected spectra. The $\alpha$ particle in direct reactions at intermediate energies is selective to the excitation of natural parities (i.e., $J^{\pi}=1^-, 2^+, ...$). The excitation cross section is measured by the singles $\alpha$-scattering cross section, $\mbox{d}\sigma/\mbox{d}\Omega_{\alpha}$. It is determined from the summed $\gamma$-ray spectrum of all HPGe detectors (see Fig. \ref{fig:GS_gamma}) obtained by gating on equal excitation energy, $E_X$, and $\gamma$-ray energy, $E_{\gamma}$, in the 2-dimensional spectra of excitation energy versus $\gamma$-ray energy (not shown). Furthermore, the HPGe-detector array allows us to measure double-differential cross sections, $\mbox{d}^2\sigma/(\mbox{d}\Omega_{\alpha}\mbox{d}\Omega_{\gamma})$, which are sensitive to the multipolarity of transitions. Details about the setup and the data analysis can be found in Refs. \cite{Savr06a,Endr09,Dery13}.

\begin{figure}[htbp]
\centering
\includegraphics[width=8cm]{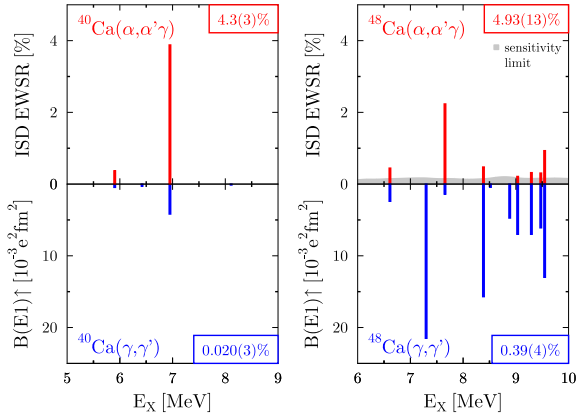}
\caption{Low-lying $E1$ strength in $^{40}$Ca \cite{Poel92,Hart00} and $^{48}$Ca (\cite{Hart00,Hart02} and present work), obtained in an $(\alpha,\alpha'\gamma)$ experiment (upper panel) and a $(\gamma,\gamma')$ experiment (lower panel). The grey-shaded area indicates the experimental sensitivity limit for the $(\alpha,\alpha'\gamma)$ experiment on $^{48}$Ca. The framed percentages give the summed exhaustion of the respective ISD (upper panel) and IV (lower panel) EWSR.}
\label{fig:ewsrbe1}
\end{figure}

In total seven $J^{\pi}=1^-$ states in $^{48}$Ca were excited by the $\alpha$ particles within the sensitivity limit, which was determined by integrating the background in the $\gamma$-ray spectrum. The excitations were identified via $\gamma$ decays to the ground state. The states at 9.47 and 9.55~MeV were observed qualitatively in previous inelastic $\alpha$-scattering experiments \cite{Pete65,Lipp67,Fuji88}. The fraction of the exhausted EWSR for isoscalar dipole (ISD) transitions (ISD EWSR) \cite{Hara81} derived from the $(\alpha,\alpha'\gamma)$ experiment is shown in Fig. \ref{fig:ewsrbe1} in comparison to $B(E1)${$\uparrow$} values from a $(\gamma,\gamma')$ experiment on $^{48}$Ca \cite{Hart00}. The singles $\alpha$-scattering cross sections were converted into fractions of the ISD EWSR by using the coupled-channels program CHUCK \cite{chuck3} and the program code BEL \cite{belgen} (with a global optical potential \cite{Nolt87} and the Harakeh-Dieperink dipole form factor \cite{Hara81}). For each $J^{\pi}=1^-$ state, the differential cross-section calculated with CHUCK was averaged over the opening angle of the BBS and compared to the experimentally determined singles $\alpha$-scattering cross section. The resulting deformation parameter allows calculating the transition rate which is needed to determine the fraction of the sum rule exhausted by a transition. In addition to $^{48}$Ca, we show the same plot for the $N=Z$ nucleus $^{40}$Ca \cite{Poel92}.

Two known isoscalar dipole excitations in $^{40}$Ca were studied in a previous $(\alpha,\alpha'\gamma)$ experiment at 120~MeV by Poelhekken \textit{et al.} \cite{Poel92}. In contrast to the experiment on $^{48}$Ca, the covered range of $\alpha$-scattering angles was between $-3^\circ \leq \theta_{lab} \leq 3^\circ$. The total $B(E1)${$\uparrow$} strength in $^{40}$Ca is mainly carried by one excitation, which belongs to the state with the largest exhaustion of the ISD EWSR. For $^{40}$Ca the $\alpha$-scattering cross sections are roughly proportional to the $E1$ strengths deduced from the $(\gamma,\gamma')$ experiment. The situation in $^{48}$Ca differs from that in $^{40}$Ca in both, the total $E1$ strength and its fragmentation into mainly seven excitations. More remarkable and surprising is the difference in the response to the isovector ($\gamma$ ray) and isoscalar ($\alpha$ particle) probes: Two excitations show a converse behavior. The $J^{\pi}=1^-$ state at 7.3~MeV, which has the strongest $E1$ transition probability in the $(\gamma,\gamma')$ experiment, was not excited by the $\alpha$ particles within an experimental sensitivity limit for the cross section of 0.15(4)~mb/sr. On the contrary, the state at 7.6~MeV, which provides one of the weakest transitions in the $(\gamma,\gamma')$ experiment, has the largest cross section for excitation by $\alpha$ particles amounting to 1.61(9)~mb/sr. We analyzed this phenomenon in detail as described below.

The spin of the observed $J=1$ states was determined in a $(\gamma,\gamma')$ experiment \cite{Hart02} by means of the angular distribution of the scattered $\gamma$ rays. For the states at 8.9 and 9.3~MeV observed in the $(\gamma,\gamma')$ experiment, $J=1$ was assigned as a result of a new reanalysis of the data. This corrects the total $B(E1)${$\uparrow$} strength to 80(8)$\cdot 10^{-3}$e$^2$fm$^2$ compared to the results published in \cite{Hart02}.

\begin{figure}[htbp]
\centering
\includegraphics[width=8cm]{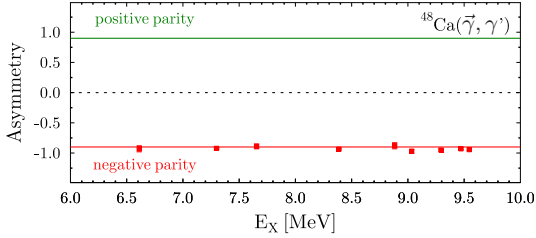}
\caption{Experimental asymmetries of $J=1$ states in $^{48}$Ca obtained in the $(\vec{\gamma},\gamma')$ experiment. The solid lines indicate the asymmetries for positive and negative parity states, respectively, which are expected with the present sensitivity of the setup. The sensitivity was deduced from the averaged absolute values of the experimental asymmetries. Uncertainties lie within the size of the markers.}
\label{fig:asymmetry}
\end{figure}

Since the $\alpha$ particle in inelastic scattering is selective to the excitation of natural parities, a possible explanation for the non-excitation of the state at 7.3~MeV could be a positive parity (i.e., $J^{\pi}=1^+$). In order to clarify the parities, an additional $(\vec{\gamma},\gamma')$ measurement with linearly polarized photons was performed at the High Intensity Gamma-ray Source (HI$\vec{\gamma}$S) facility \cite{Well09} at the Triangle Universities Nuclear Laboratory in Durham, USA. A nearly mono-energetic beam bombarded a CaCO$_3$ target of 1015 mg weight enriched in $^{48}$Ca to 90.04\%. The scattered $\gamma$ rays were detected by five HPGe detectors. Four detectors were placed perpendicular to the beam direction, thereof two parallel and two perpendicular to the polarization axis. This setup allows an unambiguous, model-independent parity determination \cite{Piet01}. Beam energies between 6.6 and 9.51~MeV were used for the investigation of nine dipole excitations in $^{48}$Ca. The experimental asymmetries $\epsilon=(I_{\parallel}-I_{\perp})/(I_{\parallel}+I_{\perp})$ of the measured photon intensities in the horizontal ($I_{\parallel}$) and vertical ($I_{\perp}$) detectors are shown in Fig. \ref{fig:asymmetry}. Electric character can be certainly assigned to all excited $J=1$ states. We conclude that the absence of the state at 7.3~MeV in the $(\alpha,\alpha'\gamma)$ experiment cannot be explained by its parity.

\begin{figure}[htbp]
\centering
\includegraphics[width=8cm]{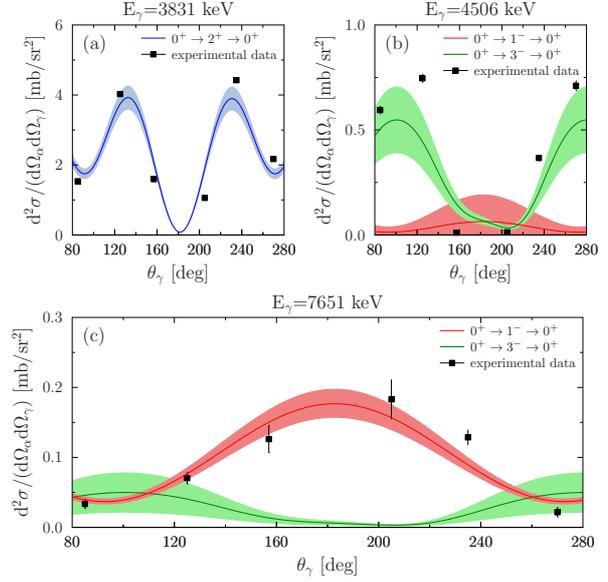}
\caption{Double-differential cross sections from the $(\alpha,\alpha'\gamma)$ experiment on $^{48}$Ca for different transitions to the ground state, together with calculated $\alpha$-$\gamma$ angular correlations for specified spin sequences. The angular correlations were fitted to the data points with one scaling parameter. The fit error is indicated by the width of the bands.}
\label{fig:ddcs}
\end{figure}

Close to the $1^-$ state at 7.6~MeV, a $3^-$ state, which can be populated in $\alpha$-scattering experiments, has previously been observed with an estimated strength of $B(E3){\uparrow}\approx 1.5$ W.u. \cite{Eise69}. Therefore, one needs to exclude a contribution of this $3^-$ state to the peak corresponding to the ground-state transition of the $1^-$ state in the $(\alpha,\alpha'\gamma)$ experiment. Ground-state transitions of this $3^-$ state are about $1000$ times weaker than ground-state transitions of the $1^-$ state.
Double-differential cross sections were determined from the $(\alpha,\alpha'\gamma)$ data and compared to $\alpha$-$\gamma$ angular distributions calculated from m-state amplitudes that were obtained from DWBA calculations \cite{chuck3,angcor} to clarify the origin of the observed ground-state transitions (see Fig. \ref{fig:ddcs}). Results for the $2^+$ state at 3.8~MeV and the first $3^-_1$ state in $^{48}$Ca at 4.5~MeV are in clear agreement with the calculated angular correlations for quadrupole and octupole transitions, respectively (see Fig. \ref{fig:ddcs}).
For the transition at 7.6~MeV, a pure dipole character is favored in a fit using a linear combination of $E1$ and $E3$ transitions (see Fig. \ref{fig:ddcs}(c)). Therefore, a strong contribution from the ground-state transition of the $3^-$ state that might affect the observed cross section of the $1^-$ state at 7.6~MeV can be excluded.
Hence, the reason for the different excitation behavior of the states at 7.3 and 7.6~MeV must lie in essentially different structures of these states. This is exposed by the different properties of the experimental probes. In addition to the isospin character, the localization and kind of interaction and the (un)selectivity to natural parities, distinguish $\gamma$ rays and $\alpha$ particles. For a complete interpretation, input from theory is needed.

\begin{figure}[htbp]
\centering
\includegraphics[width=7cm]{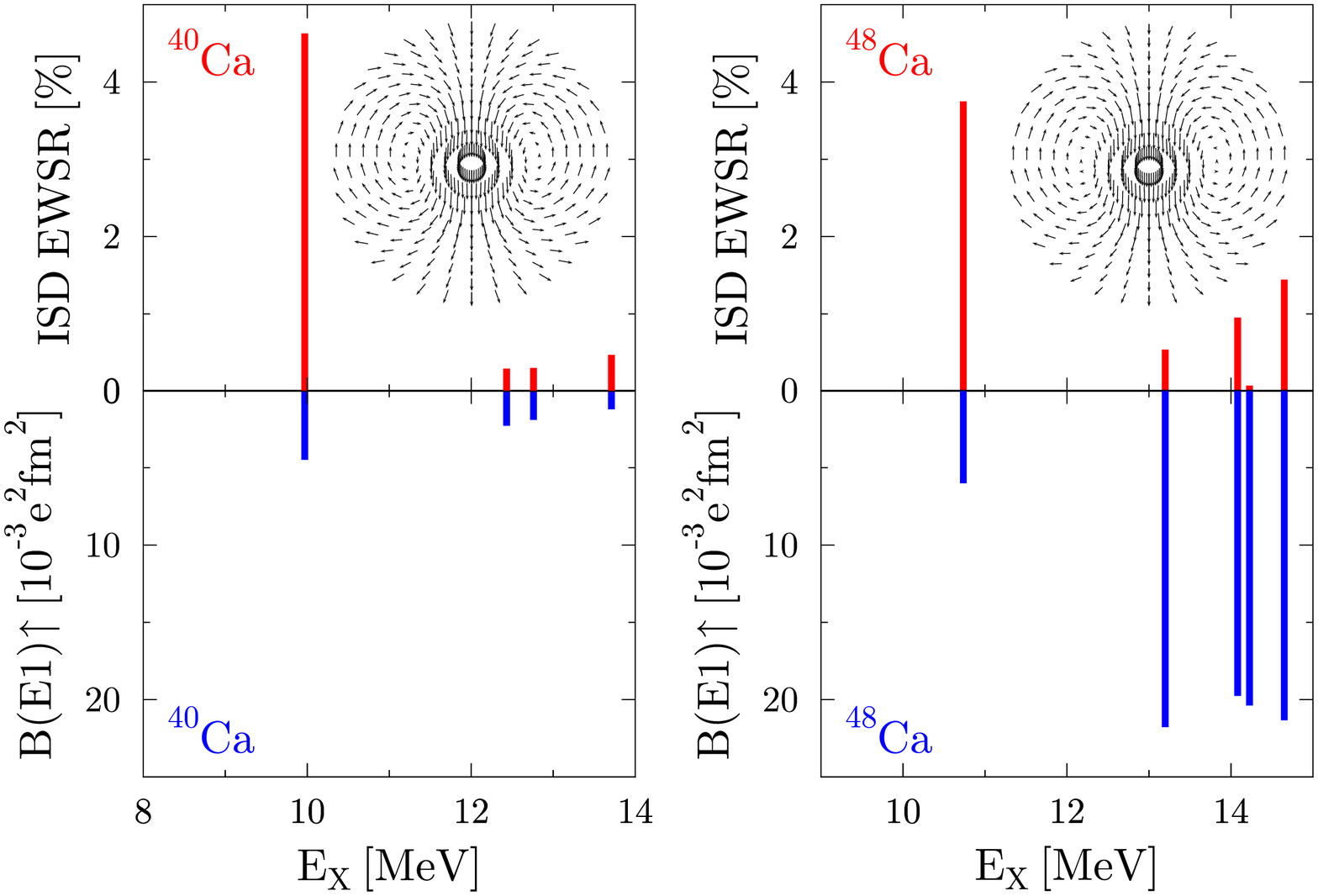}
\caption{Low-lying ISD and $B(E1){\uparrow}$ strength in $^{40}$Ca (left panel) and $^{48}$Ca (right panel) obtained from RPA calculations using the Gogny D1S interaction. These results were shown in a different way in Ref. \cite{Papa12}. In addition, IS velocity fields for the lowest-lying excitation are shown, which are presented for the first time for $^{48}$Ca.}
\label{fig:theory}
\end{figure}

Different theoretical approaches are available that, in conjunction with the experimental results, may help to clarify the nature of the $E1$ excitations in the Ca isotopes \cite{Cham94,Hart04,Tert07a,Inak11,Gamb11,Yuek12}. In an early study, Chambers \textit{et al.} \cite{Cham94} applied density functional theory and predicted a linear dependence of the low-lying $E1$ strength on the excess neutrons in the Ca chain. This behavior was not found experimentally, since the $E1$ strength up to 10~MeV in $^{44}$Ca is similar to that in $^{48}$Ca. Further calculations \cite{Tert07a} pointed out that phonon coupling (as well as coupling to the single-particle continuum, pairing, and complex configurations) is required to reproduce the $(\gamma,\gamma')$ data. Recently, the $E1$ response in even Ca isotopes was investigated along with the ISD response on basis of first-order quasiparticle RPA using the Gogny D1S interaction \cite{Papa12}. Over the whole range of neutron numbers ($N=14-40$) different scenarios explain the origin and nature of ISD strength which is present in all Ca isotopes. Proton-skin oscillation, pure IS oscillation, as well as neutron-skin oscillation are the generating mechanisms, where the latter one is predicted for $N\geq 30$. The RPA results in terms of strength distributions are shown in Fig. \ref{fig:theory} for both doubly-magic Ca isotopes. Thus, a combined comparison of the $(\gamma,\gamma')$ data with the $E1$ response and the $(\alpha,\alpha'\gamma)$ data with the ISD response is possible. An energy shift of around 3~MeV to lower energies is generally needed for reproducing the data in the Ca isotopes \cite{Papa12}. Such an energetic discrepancy with data is consistent with other RPA studies of various nuclei (see, e.g., Ref. \cite{Gori02}) and is expected to be caused by coupling to complex configurations and low-lying phonons, effects which RPA does not take into account. For lighter isotopes up to $^{48}$Ca one excitation with a dominantly IS character and almost constant ISD strength is predicted. This state is visible for $^{40,48}$Ca in Fig. \ref{fig:theory} and can be assigned to the strongly excited state in the $(\alpha,\alpha'\gamma)$ experiments on $^{40}$Ca and $^{48}$Ca at 6.9 and 7.6~MeV, respectively. The corresponding IS velocity fields (see Fig. \ref{fig:theory}) show similar flows for both nuclei, while the transition densities and form factors are similar as well \cite{Papa12}, with an oscillation of a surface layer against a core and a toroidal surface oscillation.
From this comparison of our data with theory we infer that the state at 7.6~MeV is likely an almost pure, coherent IS state and, therefore, $N=28$ marks the transition from IS to mixed, neutron-skin oscillation ($N\geq 30$). A pure IV state as observed experimentally at 7.3~MeV, is not predicted close by the strong ISD state, but at higher excitation energy. In general the stronger IV states lie above the IS state in RPA, unlike the experimental results. A likely explanation is that RPA does not take into account coupling to low-lying two-phonon states and the energetic shift of the particle-hole states close to the Fermi level, which, in the present model, are predominately responsible for the low-lying dipole strength. As in the experimental data, there is a similar IS-IV pattern for $^{40}$Ca and a different IS-IV pattern for $^{48}$Ca. These results give rise to the conjecture that the experimentally observed difference in the dipole response of $^{48}$Ca is due to different isospin characters of the excitations. In comparison to the self-conjugate nucleus $^{40}$Ca, the dipole spectrum gets richer showing IS, IV, and isospin-mixed nature which indicates a coexistence of different dipole modes with the possibility to mix. We note that the $E1$ strength of each state is much less than one single-neutron unit \cite{harakeh}. In both Figs. 2 and 5 and for both isotopes we see that much more of the total ISD strength is carried by the low-lying excitations compared to the IV strength and in all cases most of the ISD strength is carried by only one excitation, which points to the coexistence of different underlying mechanisms in the two different channels.

Isospin mixing might impact the dipole response and is therefore worth to be investigated in more detail. An optimal test case is found experimentally in $^{48}$Ca with the close-lying states at 7.3 and 7.6~MeV which are almost pure in isospin character and separated from other excitations. Therefore, a determination of the isospin-mixing matrix element becomes possible. If one assumes a two-state isospin mixing \cite{casten} of these states with initially unperturbed pure IV and IS states, a squared mixing amplitude of $\beta^2=0.061(6)$ is obtained from the ${B(E1){\uparrow}}$ ratio:
\begin{equation}
\beta^2=\frac{B(E1){\uparrow}_{II}}{B(E1){\uparrow}_{I}+B(E1){\uparrow}_{II}},
\end{equation}
where $I/II$ denotes the perturbed almost pure IV/IS state at 7.3/7.6~MeV. With $\Psi_{I,II}$ being the wave function of the perturbed state and $\Phi_{1,2}$ being the wave function of the unperturbed pure IV/IS state, we get
\begin{eqnarray}
\Psi_I&=&0.969(3)\Phi_1+0.247(12)\Phi_2,\\
\Psi_{II}&=&-0.247(12)\Phi_1+0.969(3)\Phi_2.
\end{eqnarray}
It results in an isospin-mixing matrix element of
\begin{equation}
V=\Delta E_p \sqrt{\beta^2-\beta^4}=85(3)\mbox{~keV},
\end{equation}
with a perturbed level-energy difference of $\Delta E_p=357$~keV. This value represents the first experimental indication for the isospin mixing between $1^-$ states in the energy range of the PDR. It is particularly relevant for understanding in what way isospin is broken at that excitation energy for the eventual formation of neutron-skin oscillations about an isospin symmetric core. Indeed, this isospin-mixing matrix element between far off-yrast states is already one order of magnitude larger than the one found at yrast in the odd-odd $N=Z$ nucleus $^{54}$Co \cite{Lise02} with about the same mass but even larger proton number.

In summary, the nature of low-lying $E1$ excitations in the doubly-magic nucleus $^{48}$Ca was studied using complementary experimental techniques, namely real-photon scattering with bremsstrahlung and linearly polarized photons, and the $\alpha$-$\gamma$ coincidence method. In contrast to the lightest stable calcium isotope $^{40}$Ca, a comparison of the results reveals a state-to-state difference in the underlying structure of the $1^-$ states with a dominantly isovector excitation at 7.3~MeV, a strong dominantly isoscalar excitation at 7.6~MeV and excitations with isospin-mixed character showing similar patterns in the $(\gamma,\gamma')$ and $(\alpha,\alpha'\gamma)$ experiments. This behavior is qualitatively reproduced by RPA calculations which are presented in a direct comparison. In particular, our data  show the presence of a strong IS dipole state in $^{48}$Ca, contributing very little $E1$ strength and reminiscent of the IS state of $^{40}$Ca. We conclude from these experimental and theoretical results in the heaviest stable Ca isotope, $^{48}$Ca, that diverse underlying structures contribute to the total $E1$ strength with neighboring states showing different structures with abrupt changes in their isospin character. Thus, the dipole response of $^{48}$Ca does not show an isospin splitting as observed in heavier neutron-rich nuclei \cite{Savr06a,Savr06b,Endr09,Endr10,Endr12,Dery13}, where, in general, the lowest-lying states are of mixed character, while purely IV states lie higher in energy. An isospin-mixing matrix element between $1^-$ states of $^{48}$Ca in the energy range of the PDR could be determined for the first time.

\section*{Acknowledgements}
The authors thank P. von Brentano and H. Lenske as well as many participants of the Trento workshop 2012 on ``The Nuclear Dipole Polarizability and its Impact on Nuclear Structure and Astrophysics" for discussions. We further acknowledge the support of the accelerator staff at KVI and at HI$\vec{\gamma}$S during the beam times. This work was supported by the DFG (ZI 510/4-2 and SFB 634), by the European Commission within the Sixth Framework Programme through I3-EURONS (contract No. RII3-CT-2004-506065), by the Alliance Program of the Helmholtz Association (HA216/EMMI), by the Helmholtz International Center for FAIR (HIC for FAIR), by the U.S. Department of Energy Grant No. DE-FG02-97ER41033, and by the Rare Isotope Science Project of the Institute for Basic Science funded by the Ministry of Science, ICT and Future Planning and the National Research Foundation of Korea (2013M7A1A1075766). V.D. is supported by the Bonn-Cologne Graduate School of Physics and Astronomy.


\begin{thebibliography}{50}
\expandafter\ifx\csname natexlab\endcsname\relax\def\natexlab#1{#1}\fi
\providecommand{\url}[1]{\texttt{#1}}
\providecommand{\href}[2]{#2}
\providecommand{\path}[1]{#1}
\providecommand{\DOIprefix}{doi:}
\providecommand{\ArXivprefix}{arXiv:}
\providecommand{\URLprefix}{URL: }
\providecommand{\Pubmedprefix}{pmid:}
\providecommand{\doi}[1]{\href{http://dx.doi.org/#1}{\path{#1}}}
\providecommand{\Pubmed}[1]{\href{pmid:#1}{\path{#1}}}
\providecommand{\bibinfo}[2]{#2}
\ifx\xfnm\relax \def\xfnm[#1]{\unskip,\space#1}\fi
\bibitem[{Paar et~al.(2007)}]{Paar07}
\bibinfo{author}{N.~Paar}, et~al.,
\newblock \bibinfo{journal}{Rep. Prog. Phys.} \bibinfo{volume}{70}
  (\bibinfo{year}{2007}) \bibinfo{pages}{691}.
\bibitem[{Harakeh and van~der Woude(2001)}]{harakeh}
\bibinfo{author}{M.~N. Harakeh}, \bibinfo{author}{A.~van~der Woude},
  \bibinfo{title}{{Giant Resonances}}, \bibinfo{publisher}{Oxford University
  Press, Inc.}, \bibinfo{year}{2001}.
\bibitem[{Papakonstantinou et~al.(2012)}]{Papa12}
\bibinfo{author}{P.~Papakonstantinou}, et~al.,
\newblock \bibinfo{journal}{Phys. Lett. B} \bibinfo{volume}{709}
  (\bibinfo{year}{2012}) \bibinfo{pages}{270}.
\bibitem[{Brzosko et~al.(1969)}]{Brzo69}
\bibinfo{author}{J.~S. Brzosko}, et~al.,
\newblock \bibinfo{journal}{Can. J. Phys.} \bibinfo{volume}{47}
  (\bibinfo{year}{1969}) \bibinfo{pages}{2849}.
\bibitem[{Zilges et~al.(2002)}]{Zilg02}
\bibinfo{author}{A.~Zilges}, et~al.,
\newblock \bibinfo{journal}{Phys. Lett. B} \bibinfo{volume}{542}
  (\bibinfo{year}{2002}) \bibinfo{pages}{43}.
\bibitem[{Savran et~al.(2013)Savran, Aumann, and Zilges}]{Savr13}
\bibinfo{author}{D.~Savran}, \bibinfo{author}{T.~Aumann},
  \bibinfo{author}{A.~Zilges},
\newblock \bibinfo{journal}{Prog. Part. Nucl. Phys.} \bibinfo{volume}{70}
  (\bibinfo{year}{2013}) \bibinfo{pages}{210}.
\bibitem[{Tanihata et~al.(2013)Tanihata, Savajols, and Kanungo}]{Tani13}
\bibinfo{author}{I.~Tanihata}, \bibinfo{author}{H.~Savajols},
  \bibinfo{author}{R.~Kanungo},
\newblock \bibinfo{journal}{Prog. Part. Nucl. Phys.} \bibinfo{volume}{68}
  (\bibinfo{year}{2013}) \bibinfo{pages}{215}.
\bibitem[{Repko et~al.(2013)Repko, Reinhard, Nesterenko, and Kvasil}]{Repk13}
\bibinfo{author}{A.~Repko}, \bibinfo{author}{P.-G. Reinhard},
  \bibinfo{author}{V.~O. Nesterenko}, \bibinfo{author}{J.~Kvasil},
\newblock \bibinfo{journal}{Phys. Rev. C} \bibinfo{volume}{87}
  (\bibinfo{year}{2013}) \bibinfo{pages}{024305}.
\bibitem[{Tonchev et~al.(2010)}]{Tonc10}
\bibinfo{author}{A.~P. Tonchev}, et~al.,
\newblock \bibinfo{journal}{Phys. Rev. Lett.} \bibinfo{volume}{104}
  (\bibinfo{year}{2010}) \bibinfo{pages}{072501}.
\bibitem[{Savran et~al.(2011)}]{Savr11}
\bibinfo{author}{D.~Savran}, et~al.,
\newblock \bibinfo{journal}{Phys. Rev. C} \bibinfo{volume}{84}
  (\bibinfo{year}{2011}) \bibinfo{pages}{024326}.
\bibitem[{Schwengner et~al.(2013)}]{Schw13}
\bibinfo{author}{R.~Schwengner}, et~al.,
\newblock \bibinfo{journal}{Phys. Rev. C} \bibinfo{volume}{87}
  (\bibinfo{year}{2013}) \bibinfo{pages}{024306}.
\bibitem[{Tamii et~al.(2011)}]{Tami11}
\bibinfo{author}{A.~Tamii}, et~al.,
\newblock \bibinfo{journal}{Phys. Rev. Lett.} \bibinfo{volume}{107}
  (\bibinfo{year}{2011}) \bibinfo{pages}{062502}.
\bibitem[{Savran et~al.(2006)}]{Savr06b}
\bibinfo{author}{D.~Savran}, et~al.,
\newblock \bibinfo{journal}{Phys. Rev. Lett.} \bibinfo{volume}{97}
  (\bibinfo{year}{2006}) \bibinfo{pages}{172502}.
\bibitem[{Adrich et~al.(2005)}]{Adri05}
\bibinfo{author}{P.~Adrich}, et~al.,
\newblock \bibinfo{journal}{Phys. Rev. Lett.} \bibinfo{volume}{95}
  (\bibinfo{year}{2005}) \bibinfo{pages}{132501}.
\bibitem[{Klimkiewicz et~al.(2007)}]{Klim07a}
\bibinfo{author}{A.~Klimkiewicz}, et~al.,
\newblock \bibinfo{journal}{Phys. Rev. C} \bibinfo{volume}{76}
  (\bibinfo{year}{2007}) \bibinfo{pages}{051603(R)}.
\bibitem[{Wieland et~al.(2009)}]{Wiel09}
\bibinfo{author}{O.~Wieland}, et~al.,
\newblock \bibinfo{journal}{Phys. Rev. Lett.} \bibinfo{volume}{102}
  (\bibinfo{year}{2009}) \bibinfo{pages}{092502}.
\bibitem[{Hartmann et~al.(2000)}]{Hart00}
\bibinfo{author}{T.~Hartmann}, et~al.,
\newblock \bibinfo{journal}{Phys. Rev. Lett.} \bibinfo{volume}{85}
  (\bibinfo{year}{2000}) \bibinfo{pages}{274}.
\bibitem[{Hartmann et~al.(2002)}]{Hart02}
\bibinfo{author}{T.~Hartmann}, et~al.,
\newblock \bibinfo{journal}{Phys. Rev. C} \bibinfo{volume}{65}
  (\bibinfo{year}{2002}) \bibinfo{pages}{034301}.
\bibitem[{Hartmann et~al.(2004)}]{Hart04}
\bibinfo{author}{T.~Hartmann}, et~al.,
\newblock \bibinfo{journal}{Phys. Rev. Lett.} \bibinfo{volume}{93}
  (\bibinfo{year}{2004}) \bibinfo{pages}{192501}.
\bibitem[{Isaak et~al.(2011)}]{Isaa11}
\bibinfo{author}{J.~Isaak}, et~al.,
\newblock \bibinfo{journal}{Phys. Rev. C} \bibinfo{volume}{83}
  (\bibinfo{year}{2011}) \bibinfo{pages}{034304}.
\bibitem[{Sonnabend et~al.(2011)}]{Sonn11}
\bibinfo{author}{K.~Sonnabend}, et~al.,
\newblock \bibinfo{journal}{Nucl. Instr. and Meth. A} \bibinfo{volume}{640}
  (\bibinfo{year}{2011}) \bibinfo{pages}{6}.
\bibitem[{Kamerdzhiev et~al.(2004)}]{Kame04}
\bibinfo{author}{S.~Kamerdzhiev}, et~al.,
\newblock \bibinfo{journal}{Phys. Rep.} \bibinfo{volume}{393}
  (\bibinfo{year}{2004}) \bibinfo{pages}{1}.
\bibitem[{Tertychny et~al.(2007)}]{Tert07a}
\bibinfo{author}{G.~Tertychny}, et~al.,
\newblock \bibinfo{journal}{Nucl. Phys. A} \bibinfo{volume}{788}
  (\bibinfo{year}{2007}) \bibinfo{pages}{159}.
\bibitem[{Savran et~al.(2006)}]{Savr06a}
\bibinfo{author}{D.~Savran}, et~al.,
\newblock \bibinfo{journal}{Nucl. Instr. and Meth. A} \bibinfo{volume}{564}
  (\bibinfo{year}{2006}) \bibinfo{pages}{267}.
\bibitem[{Endres et~al.(2009)}]{Endr09}
\bibinfo{author}{J.~Endres}, et~al.,
\newblock \bibinfo{journal}{Phys. Rev. C} \bibinfo{volume}{80}
  (\bibinfo{year}{2009}) \bibinfo{pages}{034302}.
\bibitem[{Endres et~al.(2010)}]{Endr10}
\bibinfo{author}{J.~Endres}, et~al.,
\newblock \bibinfo{journal}{Phys. Rev. Lett.} \bibinfo{volume}{105}
  (\bibinfo{year}{2010}) \bibinfo{pages}{212503}.
\bibitem[{Endres et~al.(2012)}]{Endr12}
\bibinfo{author}{J.~Endres}, et~al.,
\newblock \bibinfo{journal}{Phys. Rev. C} \bibinfo{volume}{85}
  (\bibinfo{year}{2012}) \bibinfo{pages}{064331}.
\bibitem[{Derya et~al.(2013)}]{Dery13}
\bibinfo{author}{V.~Derya}, et~al.,
\newblock \bibinfo{journal}{Nucl. Phys. A} \bibinfo{volume}{906}
  (\bibinfo{year}{2013}) \bibinfo{pages}{94}.
\bibitem[{Tsoneva and Lenske(2008)}]{Tson08}
\bibinfo{author}{N.~Tsoneva}, \bibinfo{author}{H.~Lenske},
\newblock \bibinfo{journal}{Phys. Rev. C} \bibinfo{volume}{77}
  (\bibinfo{year}{2008}) \bibinfo{pages}{024321}.
\bibitem[{Hagemann(1999)}]{Hage99}
\bibinfo{author}{M.~Hagemann},
\newblock \bibinfo{journal}{Nucl. Instr. and Meth. A} \bibinfo{volume}{437}
  (\bibinfo{year}{1999}) \bibinfo{pages}{459}.
\bibitem[{van~den Berg(1995)}]{Berg95}
\bibinfo{author}{A.~M. van~den Berg},
\newblock \bibinfo{journal}{Nucl. Instr. and Meth. B} \bibinfo{volume}{99}
  (\bibinfo{year}{1995}) \bibinfo{pages}{637}.
\bibitem[{Peterson(1965)}]{Pete65}
\bibinfo{author}{R.~J. Peterson},
\newblock \bibinfo{journal}{Phys. Rev.} \bibinfo{volume}{140}
  (\bibinfo{year}{1965}) \bibinfo{pages}{B1479}.
\bibitem[{Lippincott and Bernstein(1967)}]{Lipp67}
\bibinfo{author}{E.~P. Lippincott}, \bibinfo{author}{A.~M. Bernstein},
\newblock \bibinfo{journal}{Phys. Rev.} \bibinfo{volume}{163}
  (\bibinfo{year}{1967}) \bibinfo{pages}{1170}.
\bibitem[{Fujita et~al.(1988)}]{Fuji88}
\bibinfo{author}{Y.~Fujita}, et~al.,
\newblock \bibinfo{journal}{Phys. Rev. C} \bibinfo{volume}{37}
  (\bibinfo{year}{1988}) \bibinfo{pages}{45}.
\bibitem[{Harakeh and Dieperink(1981)}]{Hara81}
\bibinfo{author}{M.~N. Harakeh}, \bibinfo{author}{A.~E.~L. Dieperink},
\newblock \bibinfo{journal}{Phys. Rev. C} \bibinfo{volume}{23}
  (\bibinfo{year}{1981}) \bibinfo{pages}{2329}.
\bibitem[{Comfort and Harakeh(1979)}]{chuck3}
\bibinfo{author}{J.~R. Comfort}, \bibinfo{author}{M.~N. Harakeh},
  \bibinfo{title}{{Program CHUCK3}}, \bibinfo{year}{1979}.
  \bibinfo{note}{Modified version of CHUCK}.
\bibitem[{Harakeh(1981)}]{belgen}
\bibinfo{author}{M.~N. Harakeh}, \bibinfo{title}{Bel}, \bibinfo{year}{1981}.
  \bibinfo{note}{Internal Report KVI-77 (unpublished)}.
\bibitem[{Nolte et~al.(1987)Nolte, Machner, and Bojowald}]{Nolt87}
\bibinfo{author}{M.~Nolte}, \bibinfo{author}{H.~Machner},
  \bibinfo{author}{J.~Bojowald},
\newblock \bibinfo{journal}{Phys. Rev. C} \bibinfo{volume}{36}
  (\bibinfo{year}{1987}) \bibinfo{pages}{1312}.
\bibitem[{Poelhekken et~al.(1992)}]{Poel92}
\bibinfo{author}{T.~D. Poelhekken}, et~al.,
\newblock \bibinfo{journal}{Phys. Lett. B} \bibinfo{volume}{278}
  (\bibinfo{year}{1992}) \bibinfo{pages}{423}.
\bibitem[{Weller et~al.(2009)}]{Well09}
\bibinfo{author}{H.~R. Weller}, et~al.,
\newblock \bibinfo{journal}{Prog. Part. Nucl. Phys.} \bibinfo{volume}{62}
  (\bibinfo{year}{2009}) \bibinfo{pages}{257}.
\bibitem[{Pietralla et~al.(2001)}]{Piet01}
\bibinfo{author}{N.~Pietralla}, et~al.,
\newblock \bibinfo{journal}{Phys. Rev. Lett.} \bibinfo{volume}{88}
  (\bibinfo{year}{2001}) \bibinfo{pages}{012502}.
\bibitem[{Eisenstein et~al.(1969)}]{Eise69}
\bibinfo{author}{R.~A. Eisenstein}, et~al.,
\newblock \bibinfo{journal}{Phys. Rev.} \bibinfo{volume}{188}
  (\bibinfo{year}{1969}) \bibinfo{pages}{1815}.
\bibitem[{Harakeh and Put(1979)}]{angcor}
\bibinfo{author}{M.~N. Harakeh}, \bibinfo{author}{L.~W. Put},
  \bibinfo{title}{{ANGCOR - An angular correlation program}},
  \bibinfo{year}{1979}. \bibinfo{note}{Internal Report KVI-76 (unpublished)}.
\bibitem[{Chambers et~al.(1994)Chambers, Zaremba, Adams, and Castel}]{Cham94}
\bibinfo{author}{J.~Chambers}, \bibinfo{author}{E.~Zaremba},
  \bibinfo{author}{J.~P. Adams}, \bibinfo{author}{B.~Castel},
\newblock \bibinfo{journal}{Phys. Rev. C} \bibinfo{volume}{50}
  (\bibinfo{year}{1994}) \bibinfo{pages}{R2671}.
\bibitem[{Inakura et~al.(2011)Inakura, Nakatsukasa, and Yabana}]{Inak11}
\bibinfo{author}{T.~Inakura}, \bibinfo{author}{T.~Nakatsukasa},
  \bibinfo{author}{K.~Yabana},
\newblock \bibinfo{journal}{Phys. Rev. C} \bibinfo{volume}{84}
  (\bibinfo{year}{2011}) \bibinfo{pages}{021302}.
\bibitem[{Gambacurta et~al.(2011)Gambacurta, Grasso, and Catara}]{Gamb11}
\bibinfo{author}{D.~Gambacurta}, \bibinfo{author}{M.~Grasso},
  \bibinfo{author}{F.~Catara},
\newblock \bibinfo{journal}{Phys. Rev. C} \bibinfo{volume}{84}
  (\bibinfo{year}{2011}) \bibinfo{pages}{034301}.
\bibitem[{Y\"uksel et~al.(2012)}]{Yuek12}
\bibinfo{author}{E.~Y\"uksel}, et~al.,
\newblock \bibinfo{journal}{Nucl. Phys. A} \bibinfo{volume}{877}
  (\bibinfo{year}{2012}) \bibinfo{pages}{35}.
\bibitem[{Goriely and Khan(2002)}]{Gori02}
\bibinfo{author}{S.~Goriely}, \bibinfo{author}{E.~Khan},
\newblock \bibinfo{journal}{Nuclear Physics A} \bibinfo{volume}{706}
  (\bibinfo{year}{2002}) \bibinfo{pages}{217}.
\bibitem[{Casten(2000)}]{casten}
\bibinfo{author}{R.~F. Casten}, \bibinfo{title}{{Nuclear Structure from a
  Simple Perspective}}, \bibinfo{edition}{2} ed., \bibinfo{publisher}{Oxford
  University Press, Inc.}, \bibinfo{year}{2000}.
\bibitem[{Lisetskiy et~al.(2002)}]{Lise02}
\bibinfo{author}{A.~F. Lisetskiy}, et~al.,
\newblock \bibinfo{journal}{Phys. Rev. Lett.} \bibinfo{volume}{89}
  (\bibinfo{year}{2002}) \bibinfo{pages}{012502}.

\end{thebibliography}
\end{document}